\documentclass[12pt]{iopart}

\usepackage{xcolor}
\usepackage{graphicx}% Include figure files
\usepackage{dcolumn}% Align table columns on decimal point
\usepackage{bm}
\usepackage{bbm}
\usepackage[caption=false]{subfig}
\usepackage{amsfonts}

\makeatletter
\providecommand\underarrow@[3]{%
  \vtop{\ialign{##\crcr$\m@th\hfil#2#3\hfil$\crcr
  \noalign{\nointerlineskip\kern-0.3\baselineskip}#1#2\crcr}}}
\providecommand{\underrightarrow}{%
  \mathpalette{\underarrow@\rightarrowfill@}}
\providecommand\rightarrowfill@{\arrowfill@\relbar\relbar\rightarrow}
\providecommand\arrowfill@[4]{%
  $\m@th\thickmuskip0mu\medmuskip\thickmuskip\thinmuskip\thickmuskip
   \relax#4#1\mkern-7mu%
   \cleaders\hbox{$#4\mkern-2mu#2\mkern-2mu$}\hfill
   \mkern-7mu#3$%
}

\begin{document}

\title[Third quantization of bosonic including commuting observables]{Extending third quantization with commuting observables: a dissipative spin-boson model}

\author{Luka Medic$^{1,2}$, Anton Ram\v sak$^{1,2}$, and Toma\v z Prosen$^2$}

\address{$\phantom{}^1$ Jo{\v z}ef Stefan Institute, Jamova 39, Ljubljana, Slovenia}
\address{$\phantom{}^2$ University of Ljubljana, Faculty for Mathematics and Physics,  Jadranska
	19, Ljubljana, Slovenia}
\vspace{10pt}
%\begin{indented}
%\item[]\today
%\end{indented}

\begin{abstract}
We consider the spectral and initial value problem for the Lindblad-Gorini-Kossakowski-Sudarshan master equation describing an open quantum system of bosons and spins, where the bosonic parts of the Hamiltonian and Lindblad jump operators are quadratic and linear respectively, while the spins couple to bosons via mutually commuting spin operators. Needless to say, the spin degrees of freedom can be replaced by any set of finite-level quantum systems. 
A simple, yet non-trivial example of a single open spin-boson model is worked out in some detail.
\end{abstract}  

\section{Introduction}

Exact solutions of simple but nontrivial models describing characteristic physical phenomena are of paramount importance for understanding statistical physics in nonequilibrium. While there is an abundance of such exactly solvable nonequilibrium models in classical statistical mechanics (see, e.g., a review paper~\cite{schutz}), very few explicit analytic approaches are known in the quantum realm (e.g., review papers~\cite{prosen2015matrix,landi2021non}).

A broad route to quantum non-equilibrium physics leads via the theory of open quantum systems, especially in the many-body realm. Considering a large (many-body) quantum system, one can often describe the dynamics of its (so-called central) parts within the framework of the so-called Markov approximation, neglecting the back information flow from the rest of the system (the so-called environment) to its central part. The differential equation describing the density matrix of the central system within the {\em Markov} approximation and the {\em rotating-wave} {approximation~\cite{breuer} is the unique mathematical evolution law that preserves the hermeticity, positivity, and trace of the density matrix, called the Lindblad-Gorini-Kossakowski-Sudarshan~\cite{lindblad1976generators,gorini1976completely} equation, or Lindblad equation for short. We note that the many-body Lindblad equation provides a perfect mathematical platform for preparing engineered quantum states or quantum phases of matter within cold atom and ion trap setups~\cite{kraus2008preparation,diehl2008quantum}.

Some time ago, one of the authors developed {\em canonical formalism of quantization over operator spaces} for the diagonalization of the generator of the many-body Lindblad equation -- the so-called Liouvillian superoperator, or completely solving the Lindbladian initial value problem, for a general {\em quadratic} Hamiltonian and a set of Lindblad jump operators which are {\em linear} in canonical creation/annihilation operators.
The original proposal for fermionic systems~\cite{prosen2008third,prosen2010spectral} was later extended to bosonic systems~\cite{prosen2010quantization} (see also Ref.~\cite{prosen2015observables} for a more abstract discussion of quantization over operator spaces), and further developed by other authors~\cite{dzhioev2011super,guo2017solutions,barthel2021solving}.  Specifically, the latter reference extended the technique to include quadratic Hermitian jump operators, allowing the analytical treatment of nonequilibrium phase transitions~\cite{zhang2022criticality} and crossovers between ballistic and diffusive as well as quantum and classical transport~\cite{eisler2011crossover,temme2012stochastic}. Note, however, that even within the class of linear jump operators, one can discuss non-trivial critical 
phenomena in translationally invariant~\cite{eisert2010noise,hoening2012critical} and so-called boundary-driven systems~\cite{landi2021non,prosen2008quantum,prosen2010exact} (in the latter, the jump processes are confined to the boundaries of the system).

Experience has shown that the kinds of systems that can be treated efficiently under the closed system (2nd quantization) formulation can also be treated efficiently under the open system (3rd quantization) formulation. However, there is one type of systems with very important applications, e.g., in quantum optics, that has somehow been left out so far, namely spin-boson systems. Some aspects of integrability and exact solvability of these systems in the closed system framework have been extensively discussed in the literature (see, e.g.~\cite{braak2011integrability} for the Rabi model and \cite{skrypnyk2009integrability} for the Jaynes-Cummings and Dicke models). A recent case is the exactly solvable model of an electron in a driven harmonic oscillator with Rashba coupling~\cite{Cadez_2013,Cadez_2014}. Coupling to a thermal bath can be treated numerically~\cite{Donvil_2020}, but the analytical approach to such problems is precisely the goal of the present work. Here we essentially provide a small extension of the third quantization method that allows us to include additional degrees of freedom with finite-dimensional phase space, provided that these degrees of freedom enter the Hamiltonian and jump operators only through commuting operators.

\section{Formal solution}

The aim of this paper is to solve the following Lindblad-Gorini-Kossakowski-Sudarshan master equation for $n$ particles:
\begin{equation}
    \label{eq:masterEquation}
	\frac{\textrm{d}\rho}{\textrm{d}t} = \hat{\mathcal{L}} \rho := - i \left[H,\rho\right] + \sum_\mu \left(2 L_\mu \rho L _\mu^\dagger - \left\{L_\mu^\dagger L_\mu, \rho\right\} \right),
\end{equation}
where $H$ and $L_\mu$ are the Hamiltonian and Lindblad operators, respectively. $\rho$ is the density operator describing the state of the $n$ particles. The operators $H$ and $L_\mu$ are given by:
\begin{eqnarray}
H &=& \underline{{a}}^\dagger \cdot \mathbf{H} \underline{{a}} + \underline{{a}} \cdot \mathbf{K} \underline{{a}} + \underline{{a}}^\dagger \cdot \bar{\mathbf{K}} \underline{{a}}^\dagger + \underline{\sigma} \cdot \mathbf{\Omega} \underline{{a}} + \underline{{a}}^\dagger \cdot \mathbf{\Omega}^\dagger \underline{\sigma}, \\
L_\mu &=& \underline{l}_\mu \cdot \underline{{a}} + \underline{k}_\mu \cdot \underline{{a}}^\dagger + \underline{w}_\mu \cdot \underline{\sigma},
\end{eqnarray}
where $\underline{{a}}$ and $\underline{{a}}^\dagger$ are $n$-dimensional vectors of canonical bosonic annihilation/creation operators, and $\underline{\sigma}$ is an $n$-dimensional vector of mutually commuting Hermitian operators, with a finite discrete spectrum. We also assume that $[a_j,\sigma_{j'}]=0$. The matrix $\mathbf{H}$ is Hermitian ($\mathbf{H}^\dagger=\mathbf{H}$), the matrix $\mathbf{K}$ is symmetric ($\mathbf{K}=\mathbf{K}^T$), and the matrix $\mathbf{\Omega}$ is arbitrary. The vectors $\underline{l}_\mu$, $\underline{k}_\mu$, and $\underline{w}_\mu$ are $n$-dimensional vectors of constants.

Below we follow the notation and formalism developed in Ref. \cite{prosen2010quantization} extending $n$ bosonic degrees of freedom with additional $n$ finite level quantum systems (e.g. spins where $\sigma_j$ can be considered as their 
$z-$projections).
We introduce operators $\hat{a}_{\nu,j}$ and $\hat{a}'_{\nu,j}$, where $\nu=0,1$ and $j=1,...,n$:
\begin{eqnarray}
	\hat{a}_{0,j}=\hat{a}_{j}\strut^L, \qquad \hat{a}_{0,j}'=\hat{a}_{j}^\dagger\strut^L-\hat{a}_{j}^\dagger\strut^R, \nonumber\\
	\hat{a}_{1,j}=\hat{a}_{j}^\dagger\strut^R, \qquad \hat{a}_{1,j}'=\hat{a}_{j}\strut^R-\hat{a}_{j}\strut^L.
 \label{eq:ops}
\end{eqnarray}
Here, the superscripts $L$ and $R$ indicate the left- and right-multiplication maps acting on a vector space of operators 
\begin{equation}
\hat{b}^{L} |\rho\rangle = |b\rho\rangle,\quad
\hat{b}^{R} |\rho\rangle = 
|\rho b\rangle,\quad
\textrm{where}\quad 
b \equiv a_j,a^\dagger_j,\;\textrm{or}\; \sigma_j\,.    
\end{equation}
The operators (\ref{eq:ops}) satisfy the {\it almost-canonical} commutation relations:
\begin{equation}
	\left[\hat{a}_{\nu,j},\hat{a}'_{\mu,k}\right] = \delta_{\nu,\mu}\delta_{j,k},\quad  \left[\hat{a}_{\nu,j},\hat{a}_{\mu,k}\right] = \left[\hat{a}'_{\nu,j},\hat{a}'_{\mu,k}\right] = 0.
\end{equation}
% Along $n$-dimensional vectors of bosonic operators $\underline{\hat{a}},\underline{\hat{a}}^\dagger$ we introduced an $n$-dimensional vector of Hermitian operators $\underline{\sigma}$, where we assume a discrete spectrum. The matrix $\mathbf{H}$ is Hermitian $\mathbf{H}^\dagger=\mathbf{H}$, the matrix $\mathbf{K}$ is symmetric $\mathbf{K}=\mathbf{K}^T$, the matrix $\mathbf{\Omega}$ is arbitrary and $\underline{l}_\mu,\underline{k}_\mu, \underline{w}_\mu$ are $n$-dimensional vectors of constants.
In terms of these operators, we rewrite the Liouvillean as follows:
\begin{eqnarray}
\label{eq:liouvillean}
\kern-4em\hat{\mathcal{L}} = &-& i\hat{H}^L +i\hat{H}^R + \sum_\mu 2 \hat{L_\mu}\strut^L \hat{L_\mu^\dagger}\strut^R - \hat{L_\mu^\dagger}\strut^L \hat{L_\mu}\strut^L - \hat{L_\mu}\strut^R \hat{L_\mu^\dagger}\strut^R\\
 \kern-3em = &-& i \underline{\hat{a}}_0' \cdot \mathbf{H} \underline{\hat{a}}_0 + i \underline{\hat{a}}_1' \cdot \mathbf{H} \underline{\hat{a}}_1 + i \underline{\hat{a}}_1' \cdot \mathbf{K} \left(2\underline{\hat{a}}_0+\underline{\hat{a}}_1'\right) - i \underline{\hat{a}}_0' \cdot \bar{\mathbf{K}} \left(2\underline{\hat{a}}_1+\underline{\hat{a}}_0'\right) \nonumber \\
 \kern-4em &-& i \underline{\hat{a}}_0' \cdot \mathbf{\Omega}^\dagger \underline{\sigma}^L + i \underline{\sigma}^R \cdot \mathbf{\Omega} \underline{\hat{a}}_1' - i \left(\underline{\sigma}^L-\underline{\sigma}^R\right) \cdot \mathbf{\Omega} \underline{\hat{a}}_0 + i \underline{\hat{a}}_1 \cdot \mathbf{\Omega}^\dagger \left(\underline{\sigma}^R-\underline{\sigma}^L\right) \nonumber \\
\kern-4em &+& \underline{\hat{a}}_0' \cdot \left(\mathbf{N} - \bar{\mathbf{M}}\right) \underline{\hat{a}}_0 +\underline{\hat{a}}_1' \cdot \left(\bar{\mathbf{N}} - \mathbf{M}\right) \underline{\hat{a}}_1 +\underline{\hat{a}}_0' \cdot \left(\bar{\mathbf{L}}^T - \bar{\mathbf{L}}\right) \underline{\hat{a}}_1 +\underline{\hat{a}}_1' \cdot \left(\mathbf{L}^T - \mathbf{L}\right) \underline{\hat{a}}_0 \nonumber \\
 \kern-4em&-& \underline{\hat{a}}_0' \cdot \bar{\mathbf{L}} \underline{\hat{a}}_0 - \underline{\hat{a}}_1' \cdot \mathbf{L} \underline{\hat{a}}_1 + 2 \underline{\hat{a}}_0' \cdot \mathbf{N} \underline{\hat{a}}_1' + 2 \underline{\sigma}^L \cdot \mathbf{W} \underline{\sigma}^R - \underline{\sigma}^L \cdot \mathbf{W} \underline{\sigma}^L - \underline{\sigma}^R \cdot \mathbf{W} \underline{\sigma}^R \nonumber \\
 \kern-4em&+& \underline{\hat{a}}_0 \cdot \left(\mathbf{\bar{\mathbf{F}}-\mathbf{E}}\right)\left(\underline{\sigma}^L - \underline{\sigma}^R\right) + \underline{\hat{a}}_1 \cdot \left(\mathbf{\mathbf{F}-\bar{\mathbf{E}}}\right)\left(\underline{\sigma}^R - \underline{\sigma}^L\right) - \underline{\hat{a}}_0' \cdot \bar{\mathbf{E}}\underline{\sigma}^L - \underline{\hat{a}}_1' \cdot \mathbf{E}\underline{\sigma}^R \nonumber \\
\kern-4em &-& \underline{\hat{a}}_0' \cdot \mathbf{F} \left(\underline{\sigma}^L - 2\underline{\sigma}^R\right) - \underline{\hat{a}}_1' \cdot \bar{\mathbf{F}} \left(\underline{\sigma}^R - 2\underline{\sigma}^L\right),
\end{eqnarray}
where we define $n \times n$ matrices:
\begin{eqnarray}
\mathbf{M} &:=& \sum_\mu \underline{l}_\mu \otimes \underline{\bar{l}}_\mu = \mathbf{M}^\dagger, \quad \mathbf{N} := \sum_\mu 	\underline{k}_\mu \otimes \underline{\bar{k}}_\mu = \mathbf{N}^\dagger, \quad \mathbf{L} := \sum_\mu \underline{l}_\mu \otimes\underline{\bar{k}}_\mu,  \nonumber \\
\mathbf{W} &:=& \sum_\mu \underline{w}_\mu \otimes \underline{\bar{w}}_\mu = \mathbf{W}^\dagger, \quad \mathbf{E} := \sum_\mu \underline{l}_\mu \otimes \underline{\bar{w}}_\mu, \quad \mathbf{F} := \sum_\mu \underline{k}_\mu \otimes \underline{\bar{w}}_\mu.
\end{eqnarray}

To get rid of terms linear in $\underline{\hat{a}}$ and $\underline{\hat{a}}^\dagger$ we use a map $\underline{\hat{a}} \to \underline{\hat{a}}+\underline{s},\quad \underline{\hat{a}}' \to \underline{\hat{a}}'+\underline{s}'$ where $\underline{s},\underline{s}'$ are constant $n$-vectors to be determined.
Let us define $\hat{\underline{b}} = \left(\underline{\hat{a}},\underline{\hat{a}}'\right)^T = \left(\underline{\hat{a}}_0,\underline{\hat{a}}_1,\underline{\hat{a}}'_0,\underline{\hat{a}}'_1\right)^T$, $\underline{d} = \left(\underline{s},\underline{s}'\right)^T = \left(\underline{s}_0,\underline{s}_1,\underline{s}'_0,\underline{s}'_1\right)^T$ and $\underline{\tilde{\sigma}} = \left(\underline{\sigma}^L,\underline{\sigma}^R\right)^T$. The Liouvillean can be compactly written in a matrix form

\begin{equation}
\label{Eq:L_matrix}
\hat{\mathcal{L}} = \left(\hat{\underline{b}} - \underline{d} \right) \cdot \mathbf{S} \left(\hat{\underline{b}} - \underline{d} \right) - \hat{S}_0 
% \hat{\mathbbm{1}}
\end{equation}
where
\begin{equation}
\mathbf{S} = \begin{pmatrix}{
	\mathbf{0}      & -\mathbf{X} \cr
	-\mathbf{X}^T &  \mathbf{Y}}
\end{pmatrix},
\end{equation}
\begin{equation}
\mathbf{X} := \frac{1}{2} \begin{pmatrix}{
	i\bar{\mathbf{H}}-\bar{\mathbf{N}}+\mathbf{M} & -2i\mathbf{K}-\mathbf{L}+\mathbf{L}^T \cr
	2i\bar{\mathbf{K}}-\bar{\mathbf{L}}+\bar{\mathbf{L}}^T & -i\bar{\mathbf{H}}-\mathbf{N}+\bar{\mathbf{M}}}
\end{pmatrix},
\end{equation}
\begin{equation}
\mathbf{Y} := \frac{1}{2} \begin{pmatrix}{
	-2i\bar{\mathbf{K}}-\bar{\mathbf{L}}-\bar{\mathbf{L}}^T & 2\mathbf{N} \cr
	2\mathbf{N}^T & 2i\mathbf{K}-\mathbf{L}-\mathbf{L}^T}
\end{pmatrix} = \mathbf{Y}^T.
\end{equation}

In order to eliminate the linear terms, we must fulfill the following condition for the vector $\underline{d}$:
\begin{equation}
2\mathbf{S} \underline{d} = - \mathbf{G} \underline{\tilde{\sigma}}, \quad \textnormal{where} \quad \mathbf{G} := \begin{pmatrix}{
	\bar{\mathbf{F}}-\mathbf{E}-i\mathbf{\Omega}^T & -\bar{\mathbf{F}}+\mathbf{E}+i\mathbf{\Omega}^T \cr
	-\mathbf{F}+\bar{\mathbf{E}}-i\bar{\mathbf{\Omega}}^T & \mathbf{F}-\bar{\mathbf{E}} \cr
	-\mathbf{F}-\bar{\mathbf{E}}-i\bar{\mathbf{\Omega}}^T & 2 \mathbf{F} \cr
	2 \bar{\mathbf{F}} & -\bar{\mathbf{F}}-\mathbf{E}+i\mathbf{\Omega}^T}
\end{pmatrix}.
\end{equation}
If $\mathbf{S}$ is nonsingular, then we can compute the vector $\underline{d} = - \frac{1}{2}\mathbf{S}^{-1}\mathbf{G} \underline{\tilde{\sigma}}$. Further, we can calculate the last term in Eq. (\ref{Eq:L_matrix}):
\begin{eqnarray}
\hat{S}_0 &=& \tr\mathbf{X} - \underline{d} \cdot \mathbf{S} \underline{d} - \underline{d} \cdot \mathbf{G} \underline{\tilde{\sigma}} - \underline{\sigma}^L \cdot \mathbf{W} \left(\underline{\sigma}^R - \underline{\sigma}^L\right) - \left(\underline{\sigma}^L - \underline{\sigma}^R\right) \cdot \mathbf{W} \underline{\sigma}^R \nonumber \\
&=& \tr\mathbf{M} - \tr\mathbf{N} + \frac{1}{4} \underline{\tilde{\sigma}} \cdot \mathbf{G}^T \mathbf{S}^{-1} \mathbf{G} \underline{\tilde{\sigma}} - \left(\underline{\sigma}^L - \underline{\sigma}^R\right) \cdot \left(\mathbf{W} \underline{\sigma}^R - \bar{\mathbf{W}} \underline{\sigma}^L \right).
\end{eqnarray}

By literally following the procedure introduced in \cite{prosen2010quantization}, we successfully transform the Liouvillean into its final form:
\begin{equation}
	\hat{\mathcal{L}} = -2 \sum_{r=1}^{2n} \beta_r \hat{\zeta}_r' \hat{\zeta}_r - \hat{S}_0
\end{equation}
where
\begin{equation}
	\underline{\hat{\zeta}} = \mathbf{P} \left(\left(\underline{\hat{a}}-\underline{s}\right) - \mathbf{Z} \left(\underline{\hat{a}}' -\underline{s}' \right)\right),\qquad \underline{\hat{\zeta}}' = \mathbf{P}^{-1} \left(\underline{\hat{a}}' - \underline{s}'\right).
\end{equation}
The rapidities $\beta_r$ and the matrix $\mathbf{P}$ are obtained from the diagonalization of the matrix $\mathbf{X}$, namely:
\begin{equation}
	\mathbf{X} = \mathbf{P} \mathbf{\Delta} \mathbf{P}^{-1}, \qquad \mathbf{\Delta} = \textrm{diag} \left\{\beta_1,\dots,\beta_{2n}\right\},
\end{equation}
and the matrix $\mathbf{Z}$ is a solution of the continuous Lyapunov equation
\begin{equation}
	\mathbf{X}^T \mathbf{Z} + \mathbf{Z} \mathbf{X} = \mathbf{Y}.
\end{equation}

If all the rapidities are non-negative, $\forall \beta_r \geq 0$, one can obtain non-equilibrium steady states $| \textrm{NESS}^{\underline{s}} \rangle$ and all the corresponding decay modes $| \Theta_{\underline{m}}^{\underline{s}} \rangle$ ($\underline{m}\in\mathbb{Z}_+^{2n}$ is a multi-index) of the Liouvillean for each combination of joint eigenvalues $(\underline{s}^L, \underline{s}^R) \equiv \underline{s}$ of the hermitian operators $(\underline{\sigma}^L, \underline{\sigma}^R)$, such that
\begin{equation}
	\hat{\mathcal{L}} | \textrm{NESS}^{\underline{s}} \rangle = 0,
\end{equation}
or more precisely
\begin{equation}
	\hat{\zeta}_r | \textrm{NESS}^{\underline{s}} \rangle = 0,
\end{equation}
and 
\begin{equation}
	\hat{\mathcal{L}} | \Theta_{\underline{m}}^{\underline{s}} \rangle = \lambda_{\underline{m}} | \Theta_{\underline{m}}^{\underline{s}} \rangle,
\end{equation}
where
\begin{equation}
\label{eq:general_theta}
	| \Theta_{\underline{m}}^{\underline{s}} \rangle = \prod_r \frac{\left(\zeta_r'\right)^{m_r}}{\sqrt{m_r !}} | \textrm{NESS}^{\underline{s}} \rangle , \qquad \lambda_{\underline{m}}  = -2 \sum_r m_r \beta_r.
\end{equation}

\section{Examples}

Let us consider two simple examples using the above formalism. First, let us take $H = 0$ and $L_1 = a + z_1\sigma^z$, where $z_1\in\mathbb{C}$ and $\sigma^z$ is the Pauli matrix for the spin degree of freedom. Thus we have $\mathbf{M} = 1$, $\mathbf{W} = |z_1|^2$, and $\mathbf{E} = \bar{z}_1$ ($\mathbf{H}=\mathbf{K}=\mathbf{\Omega}=\mathbf{N}=\mathbf{L}=\mathbf{F}=0$). We compute the vector $\underline{d}$ which equals to

\begin{equation}
\underline{d} = \begin{pmatrix}{
	-z_1 \sigma\strut^L \cr
	-\bar{z}_1 \sigma\strut^R \cr
	-\bar{z}_1 \sigma\strut^L + \bar{z}_1 \sigma\strut^R \cr
	z_1 \sigma\strut^L - z_1 \sigma\strut^R}
\end{pmatrix},
\end{equation}
and from here we get $\hat{S}_0 = \tr\mathbf{M} = 1$. Combining the $\underline{b}$ and $\underline{d}$ vectors, we find that the result can be written in a simple form as
$\hat{\underline{b}} - \underline{d} = \left(
	\hat{A}\strut^L,\,
	\hat{A}^\dagger\strut^R,\,
	\hat{A}^\dagger\strut^L - \hat{A}^\dagger\strut^R,\,
	\hat{A}\strut^R - \hat{A}\strut^L
\right)^T := \left(\hat{A}_0,\,\hat{A}_1,\,\hat{A}_0',\,\hat{A}_1'\right)^T$, where we have defined $\hat{A}\strut^{L/R} = \hat{a}\strut^{L/R} + z_1\sigma\strut^{L/R}$, which are, in a sense, bosonic operators shifted by a $z_1 \sigma^z$ operator. 
%$\hat{\underline{b}} - \underline{d} = \begin{pmatrix}
%	\hat{a}\strut^L + z \sigma\strut\strut^L \\
%	\hat{a}^\dagger\strut^R + \bar{z}\sigma\strut^R \\
%	\left(\hat{a}^\dagger\strut^L + \bar{z}\sigma\strut^L\right) - \left(\hat{a}^\dagger\strut^R + \bar{z}\sigma\strut^R\right) \\
%	\left(\hat{a}\strut^R + z\sigma\strut^R\right) - \left(\hat{a}\strut^L + z\sigma\strut^L\right)
%\end{pmatrix}$

In this case $\mathbf{X}=\mathbf{\Delta}$ is already diagonal ($\mathbf{P}= I$), and the rapidities are $\beta_1 = \beta_2 = 1/2$. Moreover, since $\mathbf{Y}=0$, we also find that $\mathbf{Z}=0$. Consequently, we can conclude that $(\underline{\zeta},\underline{\zeta}')^T = \underline{\hat{b}}-\underline{d}$.

For each sector $\underline{s}=(s^L, s^R) \in \left\{-1,1\right\}^2$, we can compute the NESS:
\begin{equation}
\label{eq:ex1_ness}
	| \textrm{NESS}^{\underline{s}} \rangle = | \alpha_1 \rangle \langle \alpha_2 | \otimes |s^{L}\rangle\langle s^{R}|, 
\end{equation}
where $| \alpha_{1,2} \rangle$ with $\alpha_1 = -z_1 s^L$ and  $\alpha_2 = -z_1 s^R$ are coherent states, and $|s^{L}\rangle\langle s^{R}|$ represents the spin degree of freedom. One can also compute all the decay modes $| \Theta_{\underline{m}}^{\underline{s}} \rangle$ for each $\underline{s}$ and $\underline{m}$.

In the second example, we consider the case where $H = 0$, $L_1 = a + z_1\sigma^z$ and $L_2 = z_2 \sigma^z$, with $z_1, z_2 \in \mathbb{C}$. This modification leads to a change in $\mathbf{W}$ compared to the previous example, and we have $\mathbf{W} = |z_1|^2 + |z_2|^2$. Additionally, an extra term appears in $\hat{S}_0$, which is now given by:
\begin{equation}
\label{eq:S0}
\hat{S}_0 = \tr\mathbf{M} + |z_2|^2 \left( \sigma^L - \sigma^R \right)^2,
\end{equation}
however, the vector $\underline{d}$ remains unchanged. The extra term introduced by $L_2$ (i.e., the term involving $z_2$) in $\hat{S}_0$ contributes only to a spin-dependent dephasing effect.

\subsection{Initial condition and expectation values of spin}
% Here we would like to present a calculation of the time evolution of the expectation value of spin, $\langle \sigma^\mu (t) \rangle = \tr \left(\sigma^\mu \rho(t)\right)$ ($\mu=x,y,z$), given an initial condition. As we have already seen, the second example above was a rather trivial generalization of the first example, so here we will focus just on the first one, that is $L_1 = a + z \sigma^z$.
Here, we present a calculation of the time evolution of the expectation value of spin, $\langle \sigma^\mu (t) \rangle = \textrm{Tr} \left(\sigma^\mu \rho(t)\right)$ ($\mu=x,y,z$), given an initial condition. For this calculation, we will focus on the first example, where $L_1 = a + z_1 \sigma^z$.

We start by writing a general solution for the density matrix:
\begin{equation}
\label{eq:ex1_rhot}
	| \rho(t) \rangle = \sum_{\underline{s},\underline{m}} c_{\underline{m}}^{\underline{s}} e^{-\lambda_{\underline{m}} t} | \Theta_{\underline{m}}^{\underline{s}} \rangle
\end{equation} 
where $c_{\underline{m}}^{\underline{s}} $ are coefficients related to the initial condition $| \rho_0 \rangle = \sum_{\underline{s}} | \tilde\rho_0^{\underline{s}} \rangle \otimes |s^L\rangle\langle s^R|$, with $| \tilde\rho_0^{\underline{s}} \rangle$ being the bosonic part. To compute the coefficients $c_{\underline{m}}^{\underline{s}} $, we need to solve the following system of linear equations for $c_{\underline{m}}^{\underline{s}} $ obtained by combining Eqs. (\ref{eq:general_theta}), (\ref{eq:ex1_ness}), and (\ref{eq:ex1_rhot}):
\begin{equation}
\label{eq:ex1_linearSystem}
	| \tilde\rho_0^{\underline{s}} \rangle = \sum_{\underline{m}} c_{\underline{m}}^{\underline{s}} \,\frac{\left(\zeta_1'\right)^{m_1}}{\sqrt{m_1 !}} \frac{\left(\zeta_2'\right)^{m_2}}{\sqrt{m_2 !}} \,| -z_1 s^L \rangle \langle  -z_1 s^R |, \qquad \forall \underline{s}.
\end{equation}

We consider a scenario where the system is prepared in the initial state $| \rho_0 \rangle = |0\rangle\langle0| \otimes \frac{1}{2}(I +\sigma_x)$. In Fig. \ref{fig:coeffDiff}, we illustrate the numerical convergence of the coefficients $c_{\underline{m}}^{\underline{s}}$ for different truncations of the linear system Eq. (\ref{eq:ex1_linearSystem}), and two choices of the parameter $z_1$.
%\begin{figure}
%	\includegraphics[width=.5\textwidth]{coefficientDiffs2}
%	\caption{Convergence of the coefficients $c_{\underline{m}}^{\underline{s}}$ for different truncations ($\textrm{trunc}$) of the linear system (Eq. \ref{eq:ex1_linearSystem}) with respect to the truncation $\textrm{trunc}_0=30$.  Initial state is set to $| \rho_0 \rangle = \sum_{\underline{s}} |0\rangle\langle0| \otimes \frac{1}{2} \rho^{\underline{s}}$ and $z=1$.}
%	\label{fig:coeffDiff}
%\end{figure}
\begin{figure}
    \centering
        \subfloat[]{\includegraphics[width=0.35\columnwidth]{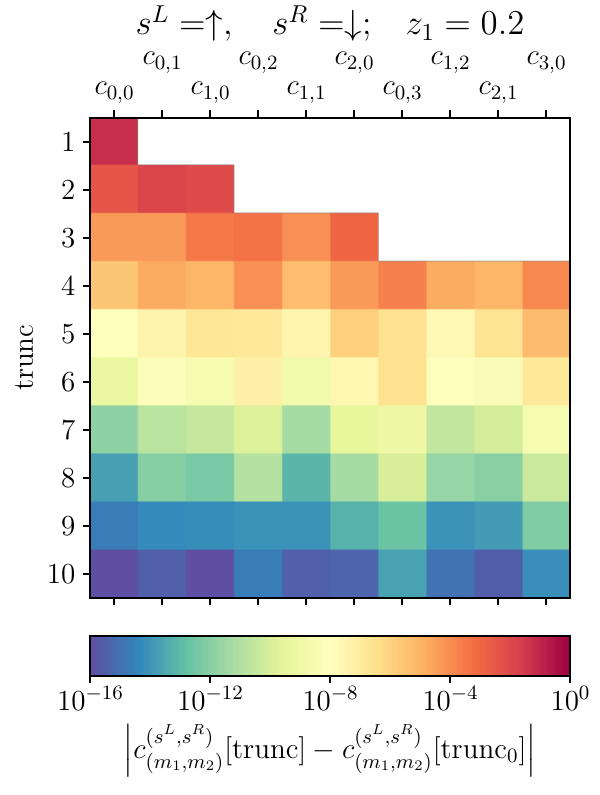}\label{fig:coeffDiff1}}
        \quad
        \subfloat[]{\includegraphics[width=0.35\columnwidth]{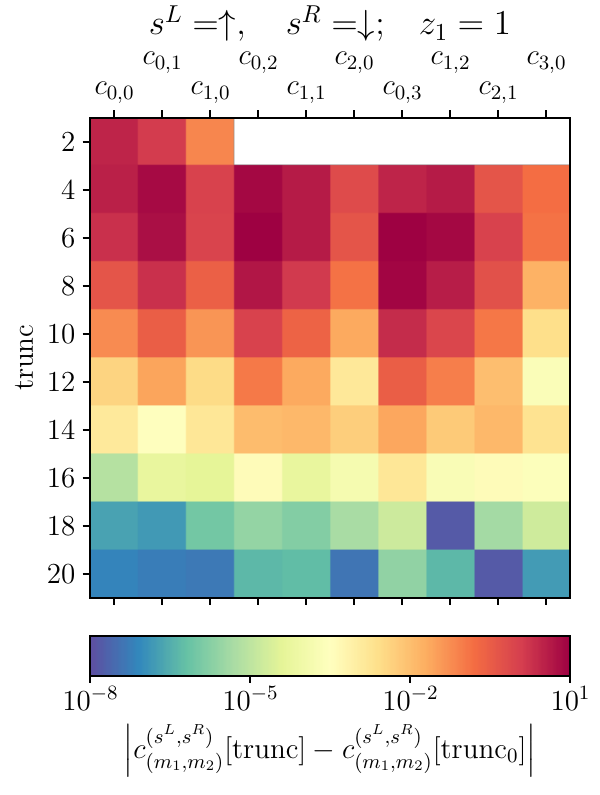}\label{fig:coeffDiff2}}
    \caption{Convergence of the coefficients $c_{\underline{m}}^{\underline{s}}$ for different truncations ($\textrm{trunc}$) of the linear system Eq. (\ref{eq:ex1_linearSystem}) with respect to the truncation $\textrm{trunc}_0=30$ for (a) $z_1=0.2$, and (b) $z_1=1$. %Initial state is set to $| \rho_0 \rangle = \sum_{\underline{s}} |0\rangle\langle0| \otimes \frac{1}{2} (I+\sigma_x)$ and $z=1$.
    }
    \label{fig:coeffDiff}
\end{figure}

Once the coefficients $c_{\underline{m}}^{\underline{s}}$ are calculated, one can compute the expectation values $\langle \sigma^\mu (t) \rangle$. For the example discussed above, the expectation values at $t=0$ are $\langle \sigma^x (0) \rangle = 1$, $\langle \sigma^y (0) \rangle = \langle \sigma^z (0) \rangle = 0$. It is evident that only the non-zero components will exhibit time-dependence, which, in this example, is $\langle \sigma^x(t) \rangle$. To test our results, we also computed $\langle \sigma^x(t) \rangle$ using the {\it Python} library {\it QuTiP} \cite{Johansson2012qutip, Johansson2013qutip2}, which allows for the computation of the time evolution of the density matrix following the Lindblad master equation. Numerical calculations for the time dependence of $\langle \sigma^x(t) \rangle$ (Fig. \ref{fig:sigmaNumeric}a) are within numerical precision fully consistent with the conjectured analytical expression:
\begin{equation}
\label{eq:sigmaXexact}
	\langle \sigma^x(t)\rangle = \langle \sigma^x(0)\rangle \exp{\left[-4|z_1|^2 \left( 1- e^{-t} \right)\right]}.
\end{equation}
In Fig. \ref{fig:sigmaNumeric}b, we illustrate the discrepancies between $\langle \sigma^x\rangle$ computed using the coefficients $c_{\underline{m}}^{\underline{s}}$ obtained from solving the linear system (\ref{eq:ex1_linearSystem}) for different truncations ($\textrm{trunc}$) and the analytical expression (\ref{eq:sigmaXexact}). As expected, the discrepancy decreases as we increase $\textrm{trunc}$. However, beyond $\textrm{trunc} \approx 18$, the solution does not improve further due to limitations in numerical precision when inverting the linear system (\ref{eq:ex1_linearSystem}) to obtain the coefficients.

%\begin{figure}
%	\includegraphics[width=0.5\textwidth]{paperPlotSigmaX}
%	\caption{Numerical solutions of the expectation value $\sigma^x$ were obtained by \texttt{python} library \texttt{qutip} (crosses), the analytical expression (\ref{eq:sigmaXexact}) matches these results (line) and the approximation (\ref{eq:sigmaXapprox}) holds for small values of $z$, e.g. $z=0.2$, and $t$ (dashed line).}
%	\label{fig:sigmaX}
%\end{figure}
%\begin{figure}
%	\includegraphics[width=0.33\textwidth]{discrepancy_3}
%	\caption{Discrepancy between ($n$) the numerical computation using the coefficients $c_{\underline{m}}^{\underline{s}}$ for different truncations ($\textrm{trunc}$)  and ($a$) the analytical solution (\ref{eq:sigmaXexact}) for the expectation value of spin $\langle \sigma^x \rangle$.}
%	\label{fig:discrepancy}
%\end{figure}
\begin{figure}
    \centering
        \subfloat[]{\quad\includegraphics[width=0.55\columnwidth]{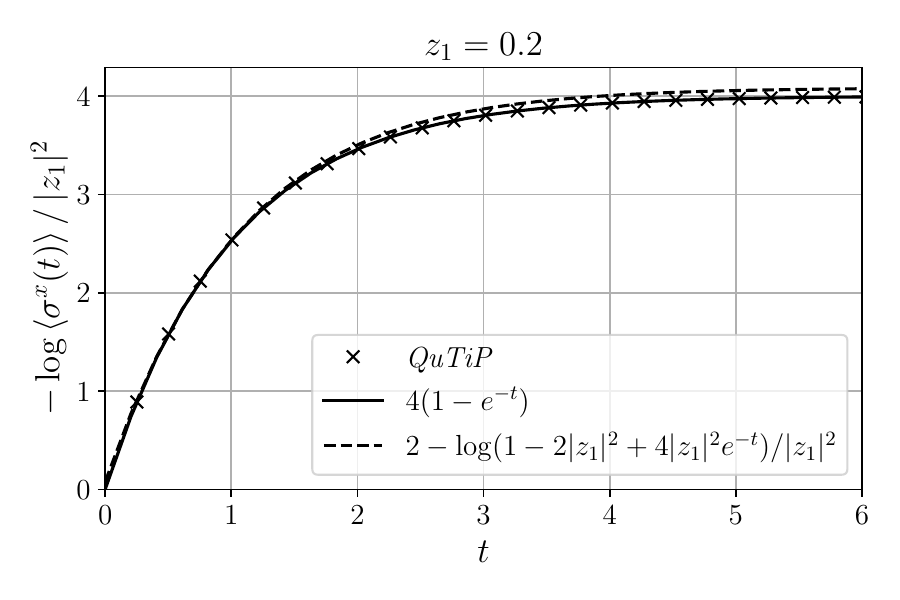}\label{fig:sigmaX}}
        \quad
        \subfloat[]{\includegraphics[width=0.37\columnwidth]{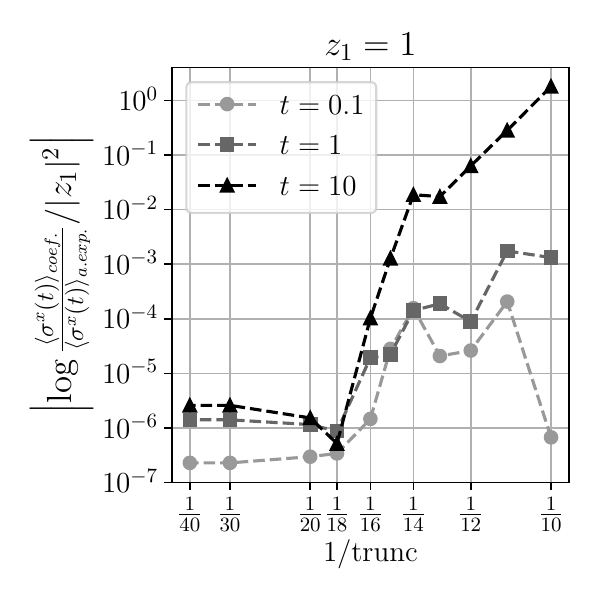}\label{fig:discrepancy}}
    \caption{Numerical and analytical solutions of the expectation value $\langle\sigma^x(t)\rangle$: (a) the numerical solution obtained by {\it QuTiP} (crosses), the analytical expression (\ref{eq:sigmaXexact}) (solid line), and the approximation (\ref{eq:sigmaXapprox}) (dashed line); (b) discrepancies between $\langle\sigma^x\rangle_{coef.}$ -- i.e. expectation value computed using the coefficients $c_{\underline{m}}^{\underline{s}}$ for different truncations ($\textrm{trunc}$) -- and the analytical expression $\langle\sigma^x(t)\rangle_{a. exp.}$ (\ref{eq:sigmaXexact}) at $t=0.1$, $1$ and $10$.}
    \label{fig:sigmaNumeric}
\end{figure}

We can analytically justify the expression in the limit $|z_1| \to 0$ (\ref{eq:sigmaXexact}). To obtain the coefficients $c_{\underline{m}}^{\underline{s}}$, we perform a Taylor expansion of the expression (\ref{eq:ex1_linearSystem}) up to quadratic terms in $z_1$. This expansion results in a $4\times 4$ system of equations to solve for each $\underline{s}=(s^L,s^R)$:
\begin{equation}
    \kern-6.3em
	\begin{pmatrix}{
	1/2 \cr
	0\cr
	0\cr
	0}
	\end{pmatrix} \kern-0.5em = \kern-0.3em e^{-|z_1|^2}  \kern-0.4em
    \begin{pmatrix}{
	1                   & z_1 s^R & z_1^* s^L & -1+|z_1|^2 s^L s^R\cr
	-z_1^* s^R         & 1-|z_1|^2 & -(z_1^*)^2 s^L s^R & z_1^* (s^L+s^R)\cr
	 -z_1 s^L            & -z_1^2 s^L s^R & 1-|z_1|^2 & z_1 (s^L+s^R)\cr
	 |z_1|^2 s^L s^R & -z_1 s^L & -z_1^* s^R & 1 - |z_1|^2 (2 + s^L s^R)}
	\end{pmatrix}  \kern-0.7em\begin{pmatrix}{
	c_{(0,0)}^{\underline{s}}\cr
	c_{(0,1)}^{\underline{s}}\cr
	c_{(1,0)}^{\underline{s}}\cr
	c_{(1,1)}^{\underline{s}}}
	\end{pmatrix}\kern-0.3em, %\qquad \forall \underline{s}=(s^L,s^R),
\end{equation}
where the solution for the coefficients is given by:
\begin{eqnarray}
\!\!\!\!\!\!\!\!\!\!\!\begin{pmatrix}{
	c_{(0,0)}^{\underline{s}}\cr
	c_{(0,1)}^{\underline{s}}\cr
	c_{(1,0)}^{\underline{s}}\cr
	c_{(1,1)}^{\underline{s}}}
	\end{pmatrix} &=& \frac{e^{|z_1|^2}}{2(1+3 |z_1|^4 +2 |z_1|^6)} \begin{pmatrix}{
	1-|z_1|^2(2-s^L s^R)+|z_1|^4 (4+2s^L s^R)\cr
	z_1^* (1-|z_1|^2) s^R\cr
	z_1\,\,\, (1-|z_1|^2) s^L\cr
	|z_1|^2(1+2|z_1|^2) s^L s^R}
	\end{pmatrix} \nonumber \\
	&\approx& \frac{1}{2} \begin{pmatrix}{
	1-|z_1|^2(1-s^L s^R)\cr
	z_1^* s^R\cr
	z_1\,\,\, s^L\cr
	|z_1|^2 s^L s^R}
	\end{pmatrix}.%,\quad \forall \underline{s}=(s^L,s^R).
\end{eqnarray}

We also need the expectation values of $\sigma^x$ for the slowest decay modes to obtain the desired result. Following the notation in \cite{prosen2010quantization}, i.e., $(A | \rho \rangle = \textrm{Tr}(A\rho)$, we have:
\begin{eqnarray}
	( \sigma^x | \Theta_{(0,0)}^{\underline{s}} \rangle &=& \delta_{(-s^L, s^R)} e^{-2|z_1|^2}, \nonumber \\
	( \sigma^x | \Theta_{(0,1)}^{\underline{s}} \rangle &=& \delta_{(-s^L, s^R)} e^{-2|z_1|^2} 2 z_1 s^R, \nonumber \\
	( \sigma^x | \Theta_{(1,0)}^{\underline{s}} \rangle &=& \delta_{(-s^L, s^R)} e^{-2|z_1|^2} 2 z_1^* s^L,\qquad \forall \underline{s},
\end{eqnarray}
where $\delta$ is the Kronecker delta. The final result in the limit $|z_1|\to 0$ for the slowest decay modes ($\lambda_{(0,0)}=0, \lambda_{(0,1)}=\lambda_{(1,0)}=-1$) is then:
\begin{equation}
\label{eq:sigmaXapprox}
	\langle \sigma^x (t)\rangle = \sum_{\underline{s},\underline{m}} c_{\underline{m}}^{\underline{s}} e^{-\lambda_{\underline{m}} t} ( \sigma^x | \Theta_{\underline{m}}^{\underline{s}} \rangle \approx e^{-2|z_1|^2} \left( 1 - 2|z_1|^2 + 4 |z_1|^2 e^{-t} \right).
\end{equation}
If we take the logarithm and approximate it for $|z_1|\to 0$, we arrive at the analytical expression (\ref{eq:sigmaXexact}). Additionally, from Fig. \ref{fig:sigmaX}, we observe that Eq. (\ref{eq:sigmaXapprox}) approximates the analytical expression (\ref{eq:sigmaXexact}) and the {\it QuTiP} solution well even for $z_1=0.2$ and $t \sim 1$.

It is essential to emphasize that the expressions (\ref{eq:sigmaXexact}) and (\ref{eq:sigmaXapprox}) are valid only for the specific initial condition $| \tilde{\rho}_0^{\underline{s}} \rangle \sim | 0 \rangle \langle 0 |$. For different initial conditions, such as $| \tilde{\rho}_0^{\underline{s}} \rangle \sim | n \rangle \langle n |$ with $n \neq 0$, one must include higher orders of $\underline{m}$ in the computation of the coefficients $c_{\underline{m}}^{\underline{s}}$ to obtain the correct result.

For completeness, let us provide the solution for the second example with $L_1 = a + z_1 \sigma^z$ and $L_2 = z_2 \sigma^z$. As mentioned before, the additional term $|z_2|^2 \left( \sigma^L - \sigma^R \right)^2$ from Eq. (\ref{eq:S0}) just adds trivially to the spin-dependent dephasing. Thus, combined with the conjectured expression (\ref{eq:sigmaXexact}), we obtain:
\begin{equation}
\label{eq:sigmaXexact2}
	\langle \sigma^x(t)\rangle = \langle \sigma^x(0)\rangle \exp{\left[-4|z_1|^2 \left( 1- e^{-t} \right) - 4 |z_2|^2\, t \right]}.
\end{equation}
In Fig. \ref{fig:sigmaX2}, we present the time evolution (\ref{eq:sigmaXexact2}) for various values of $z_2$, expressed in terms of $z_1$. Evidently, for large $t$, the solutions exhibit exponential decay with a decay rate proportional to $|z_2|^2$.

\begin{figure}
	\quad\qquad\includegraphics[width=0.65\textwidth]{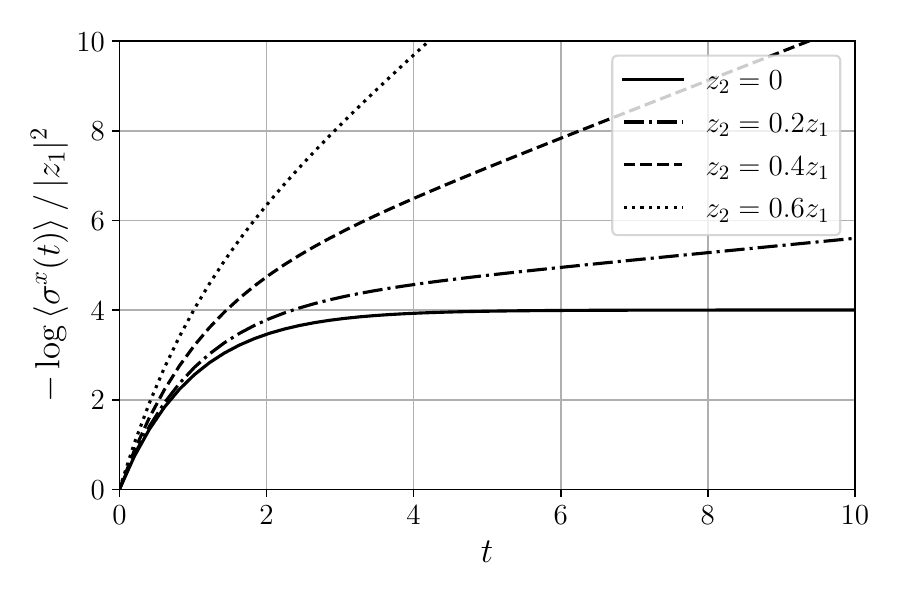}
	\caption{Time evolution of the expectation value $\langle \sigma^x (t)\rangle$ (see Eq. (\ref{eq:sigmaXexact2})) for different parameters $z_2$ and initial value $\langle \sigma^x(0) \rangle = 1$. }
	\label{fig:sigmaX2}
\end{figure}

\subsection{Limiting cases}
Lastly, let us consider two limiting cases. In the first case suppose again $|z_1|\to 0$. Then we have $\langle \sigma^x(t) \rangle \approx 1-4 |z_1|^2 (1-e^{-t})$ and in the limit $t\to\infty$ this reduces to $1-4|z_1|^2$.

In the second case, suppose slightly modified $L_1 = w a+ \sigma^z = w (a+\sigma^z / w)$ and $|w|\to 0$ such that $|w|^2 t \ll 1$ for some value of $t$. Now we get
\begin{equation}
	\langle \sigma^x (t) \rangle = \exp\left[-\frac{4}{|w|^2} (1- e^{-|w|^2 t} )\right] \underrightarrow{\phantom{}^{\phantom{}^{|w|^2 t \ll 1}}}\, e^{-4 t}
	%\xrightarrow[]{|w|^2 t \ll 1} % iopart hates amsmath
\end{equation}
and we reconstructed the exponential decoherence of the Lindbladian $L=\sigma^z$ with the rate $\gamma=2$ (note the factor 2 in the definition of the Liouvillean (\ref{eq:masterEquation})).

\section{Conclusion}
In conclusion, we have studied the dynamics of a quantum system -- which is quadratic in bosonic degrees of freedom and linear in additional commuting (classical) degrees of freedom -- in the framework of the formalism of the third quantization. In particular, we succeeded in diagonalizing the Liouville superoperator and obtaining nonequilibrium states and the corresponding decay modes for various combinations of eigenvalues of the mutually commuting Hermitian operators. A special case is the harmonic oscillator with spin degrees of freedom. Analyzing two specific examples solved exactly, we observed intriguing behavior with exponential decay rates of the spin expectation values.

For the first example, we analyzed the time evolution of the expectation value of spin $\langle \sigma^x (t) \rangle$ for $L_1 = a + z_1 \sigma^z$. We have verified numerically the dependence on the parameter $z_1$, but the analytic proof of full generality remains an open question. In the second example, we considered a modified system with the additional term $L_2 = z_2 \sigma^z$ in the Lindbladian. The analysis of this case showed an exponential decay with a decay rate proportional to $|z_2|^2$ for large $t$. The results emphasize the influence of the parameter $z_2$ on the long-term relaxation behavior of the system. Moreover, we have investigated two limiting cases and recovered the known time dependence for the Lindbladians $L=a$ and $L=\sigma_z$.

The results presented provide valuable contributions to understanding the dynamics of open quantum systems and provide a basis for further studies of related physical phenomena in which spin degrees of freedom are coupled with bosonic ones. For example, these findings could be particularly important for understanding decoherence in spin systems coupled to a bosonic thermal bath and confined in quadratic potential traps~\cite{Donvil_2020}. Such studies could provide important insights into the behavior of quantum systems under realistic experimental conditions and pave the way for advances in quantum technology and quantum information processing.

%\acknowledgements

% \newpage

\section*{Acknowledgments}
LM and AR acknowledge support from the Slovenian Research and Innovation Agency (ARIS) under Contract No. P1-0044; AR was also supported by Grant J2-2514. TP was supported by Programme P1-0402 and Grants N1-0219 and N1-0233 (ARIS).

\section*{References}
\bibliographystyle{iopart-num}
\bibliography{bibliography_list}
%\printbibliography

\end{document}